\documentstyle[aps,epsfig]{revtex}
\begin{document}
\draft
\preprint{IMSc-98/07/38; gr-qc/9807045}
\title{Black Hole Entropy and Quantum Gravity\footnote{Invited lecture
at the National Symposium on Trends and Perspectives in
Theoretical Physics, Indian Association for the Cultivation of Science,
Calcutta, April 1998.}}
\author{Parthasarathi Majumdar\footnote{email: partha@imsc.ernet.in}}
\address{The Institute of Mathematical Sciences, CIT Campus, Chennai
600113, India. }
\maketitle
\begin{abstract}
An elementary introduction is given to the problem of black hole
entropy as formulated by Bekenstein and Hawking. The information
theoretic basis of Bekenstein's formulation is briefly surveyed, and 
compared with Hawking's approach. The issue of calculating the entropy by
actual counting of microstates is taken up next, within two currently
popular approaches to quantum gravity, viz., superstring theory
and canonical quantum gravity. The treatment of the former assay is
confined to a few remarks, mainly of a critical nature, while some
computational techniques of the latter approach are elaborated. We
conclude by trying to find commonalities between these two rather
disparate directions of work.
\end{abstract}

\section{Introduction}

The intriguing possibility that the gravitational force due to a star
may be so strong that not even light could escape from it, first
appeared in Laplace's analysis \cite{lap}, almost two hundred years
ago. Using the Newtonian formula for the escape velocity of a
point mass from a gravitating sphere of mass $M$ and radius $R$, and
setting it equal to $c$, the velocity of light, Laplace obtained
the size of the gravitating sphere to be
\begin{equation}
R~~=~~{2GM \over c^2}~,\label{schw}
\end{equation}
a formula that is now well-known to yield the Schwarzschild radius of a
black hole of mass $M$. For $R~=~R_{sun}$, the radius of the
sun assumed to be a homogeneous sphere with density $\rho$, one gets
$\rho~\sim~10^{18}~ gms/cm^3~$. Matter at such a density can
hardly be stable under its self-gravity. In fact,  we now know that, after 
exhausting their nuclear fuel, stars that are still heavier than a certain
limiting mass (the Chandrasekhar mass) very likely undergo
gravitational collapse : all matter (and radiation) inside collapses to
a point, forming a spacetime singularity -- a black hole from which
nothing escapes.

While gravitational collapse {\it per se} still defies a complete
understanding, black holes have a rather precise description within
classical general relativity. They constitute a three parameter family
of exact solutions of Einstein's celebrated equation, the three
parameters being the mass $M$, the electric charge $Q$ and the angular
momentum $|{\vec L}|$. The solutions describe spacetime
geometries  with a unique point at which the (Riemann) curvature
becomes singular. In the generic case, however, the singularity is
never `naked'; it is always enshrouded by a null surface known as the
event horizon.
\begin{figure}[htb]
\begin{center}
\mbox{\epsfig{file=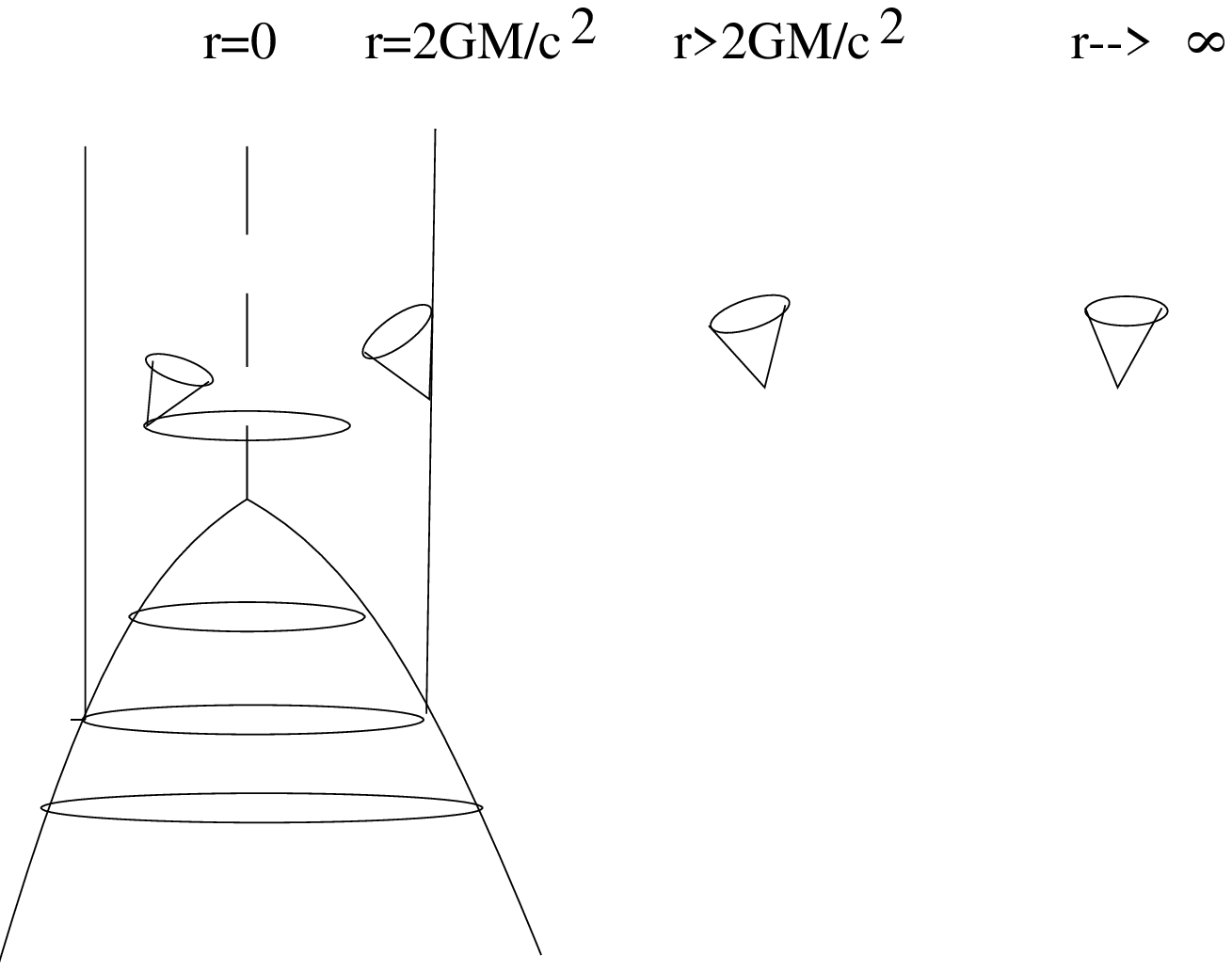,width=8truecm,angle=0}}
\caption{Gravitational collapse of a spherical star}
\end{center}
\end{figure}
Fig 1 depicts the gravitational collapse of a spherical star. One spatial
dimension has been suppressed, so that the ellipses actually represent
2-spheres at different time-slices. The envelope of the ellipses is the
spherically collapsing body. When the size of the star shrinks to that
of a sphere of
radius $R~=~2GM/c^2~\equiv~R_{Sch}$, the body is barely visible to the
external observer. This is delineated in the figure by the tipping of
the light cone as one approaches the horizon. At the horizon, the
generators of the null cone align with that surface, so that light
from the
collapsing body
grazes it. The local spacetime metric changes signature at the horizon.
As the star shrinks further, light from its surface no longer reaches
the outside
world. All null and time-like geodesics, associated with trajectories
of massless and massive particles in the black hole geometry, are
inexorably
focussed onto the curvature singularity at $r=0$. An observer on the
collapsing body, however, notices nothing special as she crosses the
event horizon.
This aspect, that the event horizon is merely a geometrical, rather
than a physical, boundary, has observational consequences \cite{nar},
but that is
another story.

Despite the fact that gravitational collapse is a cataclysmic
phenomenon wherein a multitude of physical processes (some understood,
others
yet to be discovered) are unleashed, the end-product -- a black hole
-- is a pristine object. As Chandrasekhar says, ``...the only
elements in the construction of black holes are our concepts of space and 
time. They are, thus, almost by definition, the most perfect macroscopic 
objects there are in the universe. And since the general theory of 
relativity provides a unique three-parameter family of solutions for 
their description, they are the simplest objects as well." \cite{chandra}

\section{Black Hole Entropy}

The simplicity and perfection of our conceptualization of black holes
were dramatically challenged in the early seventies by Jacob Bekenstein
\cite{bek}
and Steven Hawking \cite{haw1}, based on considerations that germinate
from the known quantum origin of all matter (and radiation). Starting
with the
simple observation that the area ${\cal A}_{hor}$ of the horizon of the
simplest black hole -- the Schwarzschild black
hole -- is a quadratic function of the mass $M$, Bekenstein \cite{bek}
noted the incremental result (in units $G=c=1$)
\begin{equation}
dM~=~\Theta~d {\cal A}_{hor}~,~~~\Theta~\equiv~1/4M~~. \label{schen}
\end{equation}
The most general black hole in general relativity, viz., the
Kerr-Newman solution, the `rationalized' area of the event horizon is
given by
\begin{equation}
{\cal A}_{hor}~=~4\pi (r_+^2~+~L^2/M^2)~,~r_{\pm}~\equiv~ M \pm
\sqrt{M^2-Q^2-L^2/M^2}~;
\label{area} \end{equation}
one then obtains similarly the incremental formula
\begin{equation}
dM~\equiv~\Theta~d {\cal A}_{hor}~+~\Phi~dQ~+~{\vec \Omega} \cdot
d{\vec L}~, ~\label{kern}
\end{equation}
where, $\Theta \equiv (r_+ - r_-)/4 {\cal A}_{hor}~,~\Phi \equiv 4\pi Q
r_+ /{\cal
A}_{hor}~,~{\vec \Omega}
\equiv 4\pi {\vec L}/M {\cal A}_{hor}~$. Eqn.s (\ref{schen}) and
(\ref{kern}) formally
resemble the First Law of thermodynamics
\begin{equation}
dU~=~TdS~+~PdV~, \end{equation}
where the second term represents the work done on the system. If we
attribute to the quantities $\Phi$ and
$\Omega$ in (\ref{kern}) above the standard interpretation of
electrostatic potential and angular velocity on
the horizon, then the second and third terms in the r.h.s. of (\ref{kern})
also represent the work done on the
black hole.

The analogy that seems to suggest itself is reinforced by Hawking's
theorem on black hole mechanics \cite{haw2}, that the
horizon area of an isolated black hole never decreases in any
transformation. E.g., if two black holes
of area ${\cal A}_1$ and ${\cal A}_2$ fuse together to form a black
hole of area ${\cal A}_{1+2}$, then the
theorem asserts that ${\cal A}_{1+2} > {\cal A}_1 + {\cal A}_2$. On the
basis of these observations and
results, Bekenstein made the bold proposal that a black hole {\it does}
have an entropy $S_{bh}$ proportional
to the area of its horizon,
\begin{equation}
S_{bh}~~=~~const.~\times~{\cal A}_{hor}~~.\label{entro}
\end{equation}
This relationship, between a thermodynamic quantity and a geometric
measure, is so striking that it warrants
an interpretation. Drawing upon Shannon's information theory
\cite{shan} and Brillouin's classic work relating
it to thermodynamics \cite{bril}, Bekenstein proposed an information
theoretic connotation for $S_{bh}$.

Consider, for instance, the isothermal compression of an ideal gas. The
thermal entropy of the gas certainly
decreases due to the compression. However, one now has better
information about the position of the molecules,
as they are now more localized. In fact, according to Brillouin, the
increase in information $\Delta I =
-\Delta S$, where, $\Delta S$ is the decrease in entropy. It follows
then that entropy measures lack of
information about the internal configurations of the system. If $p_n$
is the probability of occurrence of an
internal configuration labelled by the positive integer $n$, the
entropy is given by Boltzmann's formula (with
the Boltzmann constant $k_B =1$)
\begin{equation}
S~~=~~- \sum_n~p_n~ln p_n~. \end{equation}
The smallest unit of information is the binary bit, with $n=2$ and
$p_n=\frac12$; this corresponds to a
maximum entropy of $ln2$, which might be taken to be a unit of
entropy.

The black hole entropy $S_{bh}$ arises from our lack of information
about the nature of gravitational
collapse. The post-collapse configuration is completely characterised
by three parameters, viz.,
$M~,~Q~,~{\vec  L}$ which encode in an unknown way the diverse set of
events occurring during collapse,
just as a thermodynamic system is characterised by a few quantities
like pressure, volume, temperature
etc. which encode the microstates of the system. Thus, the black hole
entropy is not to be regarded as
the thermal entropy inside the black hole horizon. As Bekenstein
remarks, ``In fact, our black hole
entropy refers to the equivalence class of all black holes which have
the same $M~,~Q~,~L$... ."
\cite{bek}. In units where $G=c=1$, the only way in which $S_{bh}$ and
${\cal A}_{hor}$ can be
proportional is if the
constant of proportionality has the form $\eta / \hbar$, where $\eta$
is a dimensional number of $O(1)$.
The appearance of $\hbar$ is ``... a reflection of the fact that the
entropy is ... a count of states of
the system, and the  underlying states are quantum in nature ... . It
would be somewhat pretentious to
calculate the precise value of $\eta$ without a full understanding of
the quantum reality which
underlies a `classical' black hole" \cite{bek}.

The `quantum reality' Bekenstein refers to presumably subsumes quantum
gravitational effects which
inevitably occur in gravitational collapse. A complete quantum gravity
theory which serves the purpose
is still not available, although there are candidates with promise as
we discuss in the sequel. These
notwithstanding, a semiclassical estimate of $\eta$ a la' Bekenstein
may be given \cite{bek}. Eq.
(\ref{entro}) is first generalized to
\begin{equation}
S_{bh}~~=~~ f(\alpha)~, \label{entr1} \end{equation}
where, $\alpha \equiv {\cal A}_{hor} / 4\pi$ and $f$ is a monotonically
increasing function. Using
techniques of Christodolou, Bekenstein argues that the minimum increase
in the area of a black hole due
to an infalling particle of mass $\mu$ and size $b$ is given by
\begin{equation}
\left(\Delta \alpha\right )_{min}~~=~~2 \mu b_{min}~. \label{mini}
\end{equation}
Now, $b_{min} = {\lambda}_C~or~R_{sch}$, whichever is larger,
where ${\lambda}_C$ is the
Compton wavelength of the particle and $R_{sch}$, the Schwarzschild
radius; for $\mu <
(\hbar /2)^{\frac12}~,~ {\lambda}_C \ge R_{sch}$, and the other
way round for $\mu > (\hbar
/2)^{\frac12}$. In the first
case, $(\Delta \alpha)_{min} = 2 \hbar$, and in the second case,
$(\Delta \alpha)_{min} = 4 \mu^2 > 2
\hbar$. Thus, $(\Delta \alpha)_{min} = 2 \hbar$, as is indeed the case
for an `elementary' particle.
This then also quantifies the minimum loss of information due to the
particle entering the black hole
horizon. Recalling now that the minimum loss of information is a binary
bit corresponding to an
increase in entropy of $ln 2$, one sets \cite{bek}
\begin{equation}
{df \over d \alpha}~(\Delta \alpha)_{min}~~=~~ln2~,
\end{equation}
obtaining
\begin{equation}
f(\alpha)~=~\frac12{ \alpha \over \hbar}~ln2~~or~~\eta ~=~\frac12~ln2~.
\label{eta}
\end{equation}
Retrieving all factors of $G~,~c$ and $k_B$, the Bekenstein formula for
entropy, in conventional units,
is
\begin{equation}
S_{bh}~=~\frac{1}{8\pi G \hbar}~ln2~k_b c^3 {\cal A}_{hor}~~.
\label{bekent} \end{equation}
This formula is identical (except for the factor of $ln2$ which one may
think of as a choice of units
of entropy) to the one proposed by Hawking \cite{haw1} based on
consistency with the rate
of black hole radiation derived by him. We shall come to Hawking's work
shortly; prior to that two
remarks are in order, following Bekenstein.

The first of these concerns the so-called black hole temperature,
$T_{bh}$, defined in analogy with the
temperature in thermodynamics: $T^{-1} = \left({\partial S / \partial 
U}\right)_V$; here
\begin{equation}
T_{bh}^{-1}~~\equiv~~\left ({\partial S_{bh} \over {\partial
M}}\right)_{L,Q}~,
\end{equation} whence
\begin{equation}
T_{bh}~~=~~{2 \hbar \over {8\pi ln2}}~ \Theta~~, \label{temp}
\end{equation}
with $\Theta$ ($=1/4M$ for the Schwarzschild black hole) is the surface
gravity of the black hole, a
geometric quantity that remains constant over the event horizon. Once
again, Bekenstein does not favour
a {\it total} analogy with the standard notion of temperature in
thermodynamics: ``but we should
emphasize that one should not regard $T_{bh}$ as {\it the} temperature
of the black hole; such an
identification can lead to paradoxes ...".

The second remark has to do with the Second Law of thermodynamics. When
an object falls down a black
hole, there is a decrease in the thermal entropy of the universe. But,
as we have already seen, this is
invariably accompanied by an increase in the area of the horizon, with
a resultant increase in $S_{bh}$.
This led Bekenstein \cite{bek} to propose a Generalized Second Law of
thermodynamics for a universe with
black holes: $S_{thermal}~+~S_{bh}$ {\it never decreases}.

As one of his motivations for his pioneering work on black hole
radiation, Hawking considered the
situation in which a black hole (with `temperature' $T_{bh}$) is
immersed in black body radiation with
temperature $T < T_{bh}$. The question is, {\it does the black hole
radiate} as it would if it were a
thermal object at actual temperature $T_{bh}$ ? Hawking's answer is of
course an emphatic affirmative
following his discovery that black holes do emit particles in a thermal
spectrum at a temperature
$T_{bh}$ : ``... if one accepts that black holes do emit particles ...,
the identification of
$T_{bh}=\hbar \Theta /4\pi$ with the temperature of the black hole and
${\cal A}_{hor} /4 \hbar$ with the
entropy of the black hole is established, and a Generalized Second Law
confirmed ". It is clear that,
notwithstanding Bekenstein's cautionary remarks, Hawking's view of
black hole entropy (and temperature)
is in fact very close to standard thermodynamic notions.

To summarize, the black hole entropy problem consists of identifying
and counting the underlying quantum
states in an attempt to verify if the `quantum reality' actually bears
out the semiclassical
Bekenstein-Hawking formula, with the correct constant of
proportionality, in the appropriate limit. Note
that $S_{bh}$ is in fact a bit different from the standard
thermodynamic entropy which is usually a bulk
quantity, expressed as a function of the volume of the system, rather
the area of the boundary surface.
Thus, it should suffice to focus on microstates associated with the
event horizon (which, remember, is a
boundary of spacetime as seen by an external observer).

We shall not survey various semiclassical approaches to this question,
i.e., approaches not relying in a
quantum theory of gravity. These have been adequately reviewed in
another article in these proceedings
\cite{mitra}. In what follows, we consider two avenues of attack which
are claimed to be theories of
quantum gravity, namely, string theory and canonical quantum gravity.
Although the string approach to
the problem has also been nicely reviewed in another article in these
proceedings \cite{das}, we shall
make several brief remarks on that approach, some of a critical nature.
The canonical quantum gravity
approach will receive a more detailed discussion.

\section{Quantum Gravity}

A theory of quantum gravity is supposed to describe nature at a length
scale at which
quantum mechanical effects and gravitational effects become of
comparable strength.
Thus, requiring that the Compton wavelength and the Schwarzschild
radius of a particle of
mass $m$ to be of the same order, one deduces this length scale to be
of order $(G \hbar
/c^3)^{\frac12}$ -- a fundamental length scale first deduced by Planck.
Numerically, this
length is of the order $10^{-33}$ cm, corresponding to an energy of
$10^{19}$ Gev. So
far, there is no complete description of nature at such tiny lengths.
Among the
prospective candidates the most popular is (super)string theory
\cite{gsw}.

\subsection{String theory}

The basic postulate underlying this theory is that at a length scale
$l_s \ll 10^{-15} cm$,
the universe is populated by massless relativistic strings propagating
in a $D$
dimensional flat Minkowski background spacetime. Quantum mechanical
consistency of
superstring dynamics requires that $D=10$. All elementary particles of
nature are
essentially quantum excitations (modes) of the string with various
masses (starting from
zero) characterised by the string tension $\alpha'^{-1}$ where $\alpha'
\sim l_s^2$. The
massless spectrum of closed superstrings (loops) includes spin 2
gravitons, the
quanta characterising small fluctuations of spacetime geometry around
the classical
Minkowski background. Graviton scattering amplitudes in superstring
theory are believed to
be finite to all orders in string perturbation theory \cite{gsw},
although a complete
formal proof of this property is still not available.\footnote{This
finiteness of graviton
amplitudes resolves the malaise that a {\it local field} theory of
gravitons, without the
massive modes associated with strings, invariably exhibits in the
ultraviolet
\cite{deser}.} A low energy effective local field theory exclusively
for the
gravitons can be
obtained as a power series in derivatives (external momenta) from the
four-graviton
string amplitude restricted to small momenta. The first few terms in
this series are
identical, remarkably, to the first few terms in the expansion of the
Einstein-Hilbert
action around a flat Minkowski metric. However, (a) there is no
evidence that the
series derived from the string amplitude converges, and (b) even if it
does, there is no
guarantee that it will converge to the Einstein-Hilbert action.
Notwithstanding these
caveats, string theorists assume that general relativity in its
entirety is
derivable as a {\it perturbative expansion around a flat background},
and even `quantum'
corrections
to it due to virtual effects of massive string modes can be computed.
As a corollary, the
Newton constant can be
derived from string parameters : $G \sim \alpha' g_s^2$, where, $g_s$
is the
dimensionless coupling constant of superstring theory.

The hypothesis that `quantum gravity' is obtained as a perturbative expansion around a
(flat) classical background appears flawed on two counts. First of all, consider
gravitational scattering of particles at squared centre-of-mass energy $s$ and squared
momentum transfer $t$; two phenomenological dimensionless coupling parameters that appear
in the amplitudes are $Gs$ and $Gt$. The domain of quantum gravity - the Planckian regime
- is characterised by both $Gs$ and $Gt$ being of order unity or larger. In other words,
{\it quantum gravity is inherently non-perturbative}. At Planck scale, spacetime
fluctuations are anything but small, and a perturbative theory of small fluctuations can
hardly suffice to describe them, with or without the parafernelia of string theory. From a
phenomenological standpoint, unlike quantum electrodynamics or chromodynamics, quantum
`gravitodynamics' does not seem to have a weak coupling domain. At sub-Planckian energies,
Einstein's classical general relativity gives an excellent description of nature insofar
as gravitation is concerned, thereby rendering a theory of gravitons, finite or not,
physically irrelevant. Secondly, the key question of quantum gravitation is : what is the
nature of spacetime geometry at the Planck scale ? Any theory formulated in terms of a
non-dynamical classical background cannot possibly arrive at an answer, inasmuch as the
quantum theory of the hydrogen atom cannot be formulated in terms of fluctuations around
the Kepler problem for a Coulomb potential. 

Let us, however, momentarily suspend these reservations, and follow the path of string
theorists towards the issue of black hole entropy. In string theory, the assumption that
string amplitudes yield Einstein's general relativity, coupled of course to other massless
fields (dilaton, axion, Ramond-Ramond gauge fields) in the string spectrum, leads
immediately to classical black hole solutions. These solutions are characterised by a
larger parameter space than that pertaining to black holes of general relativity. In
particular, non-rotating stringy black holes\footnote{We confine our discussion here to
stringy or M-theory black holes with a unique curvature singularity and regular horizons.
Other stringy solutions with horizons that acquire curvature singularities in certain
regions of parameter space \cite{sen} fall outside the scope of this review.} carry a
number of $U(1)$ (electric-like) charges, associated with solitonic string excitations
(the so-called Ramond-Ramond states) of the $D=10$ type II superstring theory (or $D=11$
M-theory). Upon toroidal compactification of these theories down to five or four
dimensions, these appear as the so-called D(-irichlet) branes, which carry R-R charges.
Some non-rotating black holes of this theory are identified with certain (intersecting)
D-brane states. The degeneracy of these particular D-brane states can be computed exactly
in the weak coupling limit of the theory ($g_s \rightarrow 0$) as a function of the R-R
charges. In the situation that these D-brane configurations saturate the
Bogomol'nyi-Prasad-Sommerfield (BPS) bound $\sum_i Q_i = M_{bh}$, corresponding to {\it
extremal} black holes, the entropy corresponding to this degeneracy {\it matches exactly}
$S_{bh}$ as calculated from the horizon area of the black holes using the
Bekenstein-Hawking formula \cite{strv}. A clear understanding of the physics behind this
dramatic correspondence is yet to emerge. Some plausibility arguments have nevertheless
been advanced \cite{hor}, \cite{das}. 

These are based on the basic premise that the gravitational constant $G = \alpha'
g_s^2$, so that, as $g_s$ is fine tuned from weak to strong coupling, one passes from
ths string `phase' to the black hole `phase' for fixed $\alpha'$. Alternatively, the
ratio of the `Schwarzschild radius' of the string to its length varies from small to
O(1) as $g_s$ is tuned. The degeneracy formula for BPS states is argued to remain
intact during the tuning of $g_s$, appealing to non-renormalization theorems of
unbroken extended ($N > 1$) spacetime supersymmetry. Consequently, the entropy
calculated from the degeneracy appears to agree with that of the black hole
calculated from the semiclassical formula.\footnote{Ironically, the
Bekenstein-Hawking formula is known to break down in semiclassical general relativity
in the extremal situation in which the entropy of the black hole is argued to vanish
\cite{teit}, \cite{ddr}. In this case, the extremal limit is the same as exact
extremality. For stringy black holes exact extremality also yields a null entropy,
following arguments of \cite{teit}, while the extremal {\it limit} seems to agree
with the Bekenstein-Hawking formula. Clearly this warrants a better understanding.}

A major shortcoming of the foregoing correspondence is its crucial dependence on
unbroken supersymmetry, a property certainly untenable in the real world. What
happens to the degeneracy formula when supersymmetry is broken and BPS saturation is
no longer valid as a property that survives quantum corrections, is an open question.
Thus, it is a correspondence that works for a highly idealized situation, for a
special kind of black holes, most likely very different from black holes seen in
nature. The latter most likely would correspond to generic non-extremal general
relativistic black holes characterised by a far smaller parameter space than that
appearing in string theory. This is related to the necessary existence, in string
theory, of spacetime dimensions beyond the four observed in the real world. A further
technical problem is the computation of the degeneracy itself; here the practice is
to assume that the complicated intersecting D-brane configuration collapses to a
`long effective string' \cite{dm} as the entropically favoured configuration, and
then use the machinery of the 2d conformal field theory corresponding to this string,
with central charge $c=6$. The trouble with this is that it is not always
straightforward to derive this value of the central charge without extra ad hoc
assumptions. This is the situation, e.g., for four dimensional black holes obtained
in type IIB superstring theory as intersecting 2-5-6 branes \cite{hor}, where the
`long string' is non-trivial to identify.\footnote{I thank S. Das and A. Dasgupta for
a discussion on this point.} The area law would still emerge in this case, but only
upto an overall constant.

\subsection{Canonical Quantum Gravity}

This approach, also called Quantum General Relativity, envisages an exact 
solution to the problem of quantization of standard four dimensional general 
relativity, in  contrast to the previous perturbative path around a flat 
classical background. In this respect, it is closer to the canonical 
quantum theory of the hydrogen atom, for instance. 

The canonical treatment of classical general relativity, otherwise known 
as geometrodynamics \cite{adm}, is traditionally formulated in terms 
of 3-metrics, i.e, restrictions of the metric tensor to three dimensional 
spacelike hypersurfaces (`time slices'). Canonically conjugate variables 
to these are then constructed, Poisson brackets between them defined and 
the entire set of first class constraints derived. The problem with this 
is that the constraints remain quite intractable. 

A significant 
departure from this approach is to formulate canonical general 
relativity as a theory of `gauge' connections, rather than 3-metrics 
\cite{ash1}. Some of the constraints simplify markedly as a 
consequence, allowing exact treatment, although this is not true for 
all the constraints (e.g., the Hamiltonian constraint still remains 
unsolvable). The method has also undergone substantial evolution since 
its inception; a complex one-parameter family of connection variables 
is available as one's choice of the basic `coordinate' degrees of 
freedom. The original Ashtekar choice \cite{ash1}, viz., the self-dual  
$SL(2,C)$  connection (inspired by work of Amitabha Sen \cite{titu}),  
corresponding to one member of this family, is `geometrically and 
physically well-motivated' because the full tangent space group then 
becomes the gauge group of the canonical theory \cite{imm}. However, 
quantizing a theory with complex configuration degrees of freedom 
necessitates the imposition of subsidiary `reality' conditions on the 
Hilbert space, rendering the formulation unwieldy. 

A better alternative, related to the former by canonical transformations, is
to deal with the Barbero-Immirzi family of {\it real} $SU(2)$ connections confined to
the time-slice M 
\begin{equation} 
A_i^{(\beta) a} ~~\equiv~~ \epsilon
^{abc} \Omega_i~_{bc} ~ +~ \beta g_{ij} \Omega^j~_{0a}~, \label{imcon}
\end{equation} 
labelled by a positive real number $\beta (\sim O(1))$ known
as the Barbero-Immirzi (BI) parameter \cite{barb}, \cite{imm}.\footnote{Here 
$\Omega_{\mu}~^{BC}$ is the standard Levi-Civita connection, $i,j = 1,2,3$ are spatial
world indices, and $a,b,c =1,2,3$ are spatial tangent space indices, and
$g_{ij}$ is the 3-metric.} This yields the BI family of curvatures
(restricted to M), 
\begin{equation} F^{(\beta) a}_{ij}
~~\equiv~~\partial_{[i} A_{j]}^{(\beta) a}~+~\epsilon^{abc} ~A_i^{(\beta) b}~
A_j^{(\beta) c}~. \label{curv} \end{equation} 
The variables canonically
conjugate to these are given by the so-called solder form 
\begin{equation}
E_i^{(\beta) a}~~=~~\frac{1}{\beta}~\epsilon^{abc} \epsilon_{ijk} e_j~^b
e_k~^c~.  \label{sold} \end{equation} 
The canonical Poisson bracket is then given by 
\begin{equation} 
\left \{ A_i^{(\beta) a} (x)~,~E_j^{(\beta) b}(y)
\right \}~=~\beta~ \delta_{ij}~\delta^{ab}~ \delta(x,y)~. \label{pb}
\end{equation}

Canonical quantization in the connection representation implies that 
physical states are gauge invariant functionals of $A_i^{(\beta)a}(x)$ and 
\begin{equation}
E_i^{(\beta) a}~\rightarrow ~ {\hat E}_i^{(\beta) a}~\equiv~{{\beta \hbar} \over 
i}~{\delta \over {\delta A_a^{(\beta) i}}}~. \label{mom} \end{equation}
A useful basis of states for solution of the quantum constraints are the `spin 
network' states which generalize the loop space states used earlier 
\cite{rov1}. A spin network consists of a collection {\it edges} and {\it 
vertices}, such that, if two distinct edges meet, they do so in a vertex. It is 
a lattice whose edges need not be rectangular, and indeed may be non-trivially 
knotted. E.g., the graph shown in fig. 2 has 9 edges and 6 vertices. 
\begin{figure}[htb]
\begin{center}
\mbox{\epsfig{file=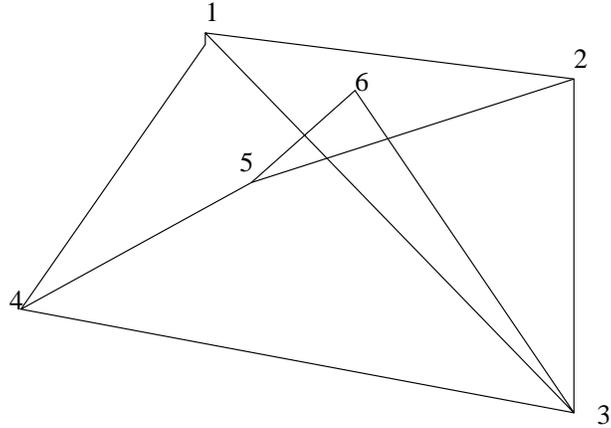,width=8truecm,angle=0}}
\caption{A spin network with 9 edges and 6 vertices}
\end{center}
\end{figure}

To every edge $\gamma_l$ ($l=1,2, ..., 9$) we assign a spin $J_l$ which takes 
all half-integral values including zero. Thus, each edge transforms as a 
finite dimensional irreducible representation of $SU(2)$. In addition, one 
assigns to each edge a Wilson line functional of the gauge 
connection $h_l(A) = {\cal P}~\exp~\int_l~d\gamma^i_l~(A \cdot \tau)_i$, where 
$\tau^a$ are $SU(2)$ generators in the adjoint representation. To every vertex 
is assigned an $SU(2)$ invariant tensor $C^v$. These assignments completely 
define the basis states, which form a dense set in the Hilbert space of gauge 
invariant functionals of ${^{\beta}}A$. The inner product of these states then 
induces a measure on the space of connections which can be used to define a 
`loop transform' \cite{imm} of physical states, representing the same state, by 
diffeomorphism invariance. `Weave' states, supported on complicated and fine 
meshed nets  (with meshes of Planck scale size) are supposedly typical physical 
states. Thus, the classical spacetime continuum metamorphoses in the quantum 
domain into a space 
of `weaves' with meshes of Planck scale size on which all curvature (and indeed 
all dynamics) is concentrated. The Einsteinian continuum emerges when we view 
the weaves from afar, and are no longer able to see the meshes. 

Observables on the space of physical states (like the weaves) include 
geometrical operators like the area and volume operators, which typically are 
functionals of the canonical variables. To calculate the spectrum of these 
operators in the connection representation requires a technique of 
`regularization' since the classical definition of these quantities translates 
into singular objects upon naive quantization. E.g., the area operator
${\hat {\cal A}} (S)$ corresponding to a two dimensional surface $S$ 
intersecting a subset ${\cal L}$ of edges of a net, not touching any of its 
vertices and having no edge lying {\it on} $S$ is formally defined as 
\begin{equation}
{\hat {\cal A}}(S)~\psi_n~\equiv~ \left( ~\int d^2 \sigma \sqrt{n_i n_j {\hat 
E}^{ia} {\hat E}^{jb}}~ \right)_{reg}~ \psi_n~. \end{equation}
For large areas compared to $l_{Planck}^2$, this reduces to \cite{rov2}, 
\cite{ash3} 
\begin{equation}
{\hat {\cal A}}(S)~\psi_n~=~\beta \hbar l_{Planck}^2 ~\sum_{l \epsilon 
{\cal L}} \sqrt{J_l (J_l+1)}~\psi_n~~. \label{darea} \end{equation}
The discreteness in the eigen-spectrum of the area operator is of course 
reminiscent of 
discrete energy spectra associated with stationary states of familiar quantum 
systems. Each 
element  of the discrete set in (\ref{darea}) corresponds to a particular {\it 
number} of  intersections (`punctures') of the spin net with the boundary surface 
$S$. Diffeomorphism invariance ensures the irrelevance of the locations of 
punctures. This will have important ramifications later.

We now consider the application of the foregoing formalism to the calculation
of entropy of the four dimensional Schwarzschild black hole, following
\cite{rov2}, \cite{car1}, \cite{kras}, \cite{ash4} and \cite{km}. The basic idea is to
concentrate on the horizon as a boundary surface of spacetime (the rest of
the boundary being described by the asymptotic null infinities ${\cal
I}^{\pm}$), on which are to be imposed boundary conditions specific to the
horizon geometry of the Schwarzschild black hole vis-a-vis its symmetries
etc. These boundary conditions then imply a certain description for the
quantum degrees of freedom on the boundary. The entropy is calculated by
counting the `number' of boundary degrees of freedom. The region of spacetime
useful for our purpose is depicted in the Penrose and Finkelstein diagrams in
fig. 3. The four-fold ${\cal M}$ has as a boundary the event horizon 
${\cal H}$ in addition to ${\cal I}^{\pm}$. $\Delta$ is a `finite patch' on 
${\cal H}$ of constant cross-sectional area ${\cal A}_S$. M is a particular 
time-slice which intersects ${\cal H}$ (in particular $\Delta$) in the 
2-sphere $S$. 
\begin{figure}[htb]
\begin{center}
\mbox{\epsfig{file=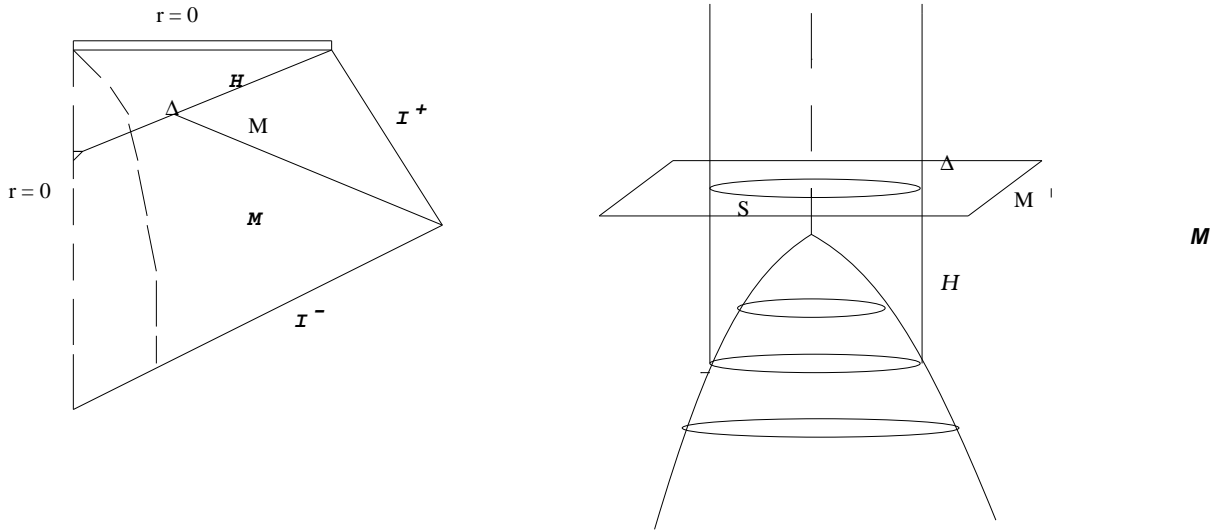,width=16truecm,angle=0}}
\caption{Penrose and Finkelstein diagrams showing the Schwarzschild black hole in 
the canonical framework} 
\end{center}
\end{figure}

Standard asymptotically flat boundary conditions are imposed on ${\cal 
I}^{\pm}$; those on the event horizon essentially subsume the following: 
first of all, the horizon is a null surface with respect to the 
Schwarzschild metric; second, the black hole is an isolated one with no 
gravitational radiation on the horizon; thirdly, the patch $\Delta$ has two flat 
(angular) coordinates spanning a special 2-sphere which {\it coincides} with 
$S$, the intersection of the time-slice $M$ with $\Delta$. The last 
requirement follows from the spherical symmetry of the Schwarzschild 
geometry. These boundary conditions have a crucial effect on the 
classical Hamiltonian 
structure of the theory, in that, in addition to the bulk contribution to 
the area tensor of phase space (the symplectic structure) arising in 
canonical general relativity, there is a {\it boundary} contribution. Notice 
that the boundary of the spacelike hypersurface M intersecting the black 
hole horizon is the 2-sphere $S$. Thus, the symplectic structure is given by
\begin{eqnarray}
\Omega|_{ A^{(\beta)}, E^{(\beta)} } (~ \delta E^{(\beta)}~,~\delta 
A^{(\beta)}~&;&~\delta E^{(\beta)'}~,~ \delta A^{(\beta)'} ~) 
\nonumber \\
&=&{1 \over {8\pi G}}~\int_{\em M}~Tr~( \delta E^{(\beta)} \wedge \delta 
A^{(\beta)'}~-~\delta E^{(\beta)'} \wedge \delta  A^{(\beta)} )~\nonumber \\
&-& {k \over {2\pi}}~\int_{S=\partial {\em M}}~ Tr ( ~\delta 
A^{(\beta)} \wedge \delta A^{(\beta)'} )~, \label{symp} \end{eqnarray}
where, $k \equiv {{\cal A}_S \over {2 \pi \beta G}}$. The second term in 
(\ref{symp}) corresponds to the boundary contribution to the symplectic 
structure; it is nothing but the symplectic structure of an $SU(2)$ level 
$k$ Chern Simons theory living on M. This is consistent with an extra term 
that arises due to the boundary conditions in the action, that is exactly an 
$SU(2)$ level $k$ Chern Simons action on the three dimensional piece 
$\Delta$ of the event horizon \cite{ash4}. As a consequence of the boundary 
Chern Simons 
term, the curvature pulled back to $S$ is proportional to the pullback (to $S$) of the 
solder form 
\begin{equation}
F^{(\beta)}~+~{2\pi \beta \over {\cal A}_S}~E^{(\beta)}~=~0~~.\label{cons} 
\end{equation}
This is a key relation for the entropy computation as we now proceed to demonstrate.

In the quantum theory, we have already seen that spacetime `in the bulk' is
described by spin nets ($\{ \psi_V \}$ say) at fixed time-slices. It has been shown
\cite{ash3} that spin network states constitute an eigen-basis for the solder form
with a discrete spectrum. Now, in our case, because of the existence of the event
horizon which forms a boundary of spacetime, there are additional surface states
$\{ \psi_S \}$ associated with Chern Simons theory. In the canonical framework, the
surface of interest is the 2-sphere $S$ which forms the boundary of M.  Thus,
typically a state vector in the Hilbert space {\bf H} would consist of tensor
product states $\psi_V \otimes \psi_S$. Eq. (\ref{cons}) would now act on such
states as an operator equation. It follows that the surface states $\{ \psi_S \}$
would constitute an eigenbasis for $F^{(\beta)}$ restricted to $S$, with a discrete
spectrum. In other words, {\it the curvature has a support on $S$ only at a
discrete set of points -- punctures}. These punctures are exactly the points on $S$
which are intersected by edges of spin network `bulk' states, in the manner
discussed earlier for the definition of the area operator. At each puncture $p$
therefore one has a specific spin $J_p$ corresponding to the edge which pierces $S$
at $p$. The black hole can then be depicted (in an approximate sense) as shown in
fig. 4.  
\begin{figure}[htb] 
\begin{center}
\mbox{\epsfig{file=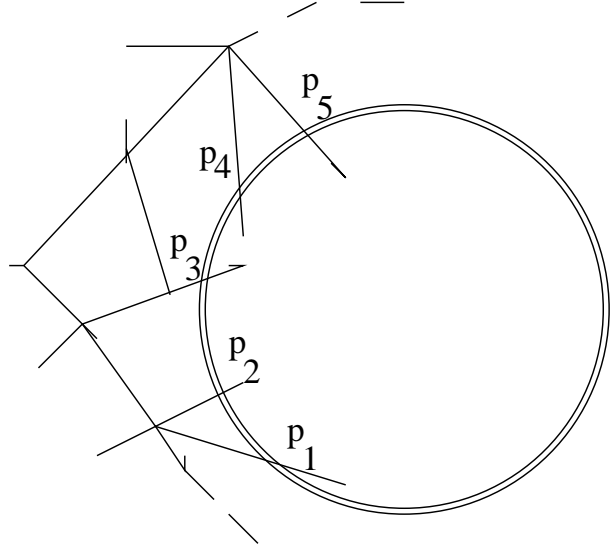,width=8truecm,angle=0}} 
\caption{The Schwarzschild black hole in the spin network picture, showing five of 
the punctures $p_1, \dots, p_5$} 
\end{center} 
\end{figure}

Consider now a set of punctures ${\cal P}_{(n)} = \{ p_1,J_{p_1}~;~p_2, J_{P_2}~;~
\dots p_n, J_{p_n} \}$.  For every such set, there is a subspace ${\bf H}_V^{\cal P}
$ of ${\bf H}_V$ which describes the space of spin net states corresponding to the
punctures. Similarly, there is a subspace ${\bf H}S^{\cal P}$ of ${\bf H}_S$
describing the boundary Chern Simons states corresponding to the punctures in
${\cal P}$. The full Hilbert space is given by the direct sum, over all possible
sets of punctures, of the direct product of these two Hilbert (sub)spaces, {\it
modulo} internal gauge transformations and diffeomorphisms.\footnote{The latter
symmetry, in particular, as already mentioned, implies that the location of punctures
on $S$ cannot have any physical significance.} Now, given that the Hamiltonian
constraint cannot be solved exactly, one assumes that there is at least one
solution of the operator equation acting on the full Hilbert space, for a given set
of punctures ${\cal P}$. 

One now assumes that it is only the surface states $\psi_S$ that 
constitute the microstates contributing to the entropy of the black hole $S_{bh}$, so 
that the volume states $\psi_V$ are traced over, to yield the black hole entropy as
\begin{equation}
S_{bh}~~=~~ln \sum_{\cal P}~dim~ {\bf H}_S^{\cal P}~~. \label{entr}
\end{equation}
The task has thus been reduced to computing the number of $SU(2)_k$ Chern Simons
boundary states for a surface with an area that is ${\cal A}_S$ to within
$O(l_{Planck}^2)$. One now recalls a well-known correspondence between the
dimensionality of the Hilbert space of the Chern Simons theory and the number of
conformal blocks of the two dimensional conformal field theory (in this case
$SU(2)_k$ Wess-Zumino-Witten model) `living' on the boundary \cite{wit}. This
correspondence now simplifies the problem further to the computation of the number
of conformal blocks of the WZW model. Thus, {\it the problem of counting the
microstates contributing to the entropy of a 4d Schwarzschild black hole has
metamorphosed into counting the number of conformal blocks for a particular 2d
conformal field theory.}

This number
can be computed in terms of the so-called fusion  matrices $N_{ij}^{~~r}$ \cite{dms} 
\begin{equation}
N^{\cal P}~=~~\sum_{\{r_i\}}~N_{j_1 j_2}^{~~~~r_1}~ N_{r_1 j_3}^{~~~~r_2}~ N_{r_2
j_4}^{~~~~r_3}~\dots  \dots~ N_{r_{p-2} 
j_{p-1}}^{~~~~~~~~j_p} ~ \label{fun} \end{equation}
This is very similar to the composition of angular momentum in ordinary quantum 
mechanics; 
it has been extended here to the infinite dimensional affine Lie algebra $SU(2)_k$. 
Diagrammatically, this can be represented as shown in fig. 5 below.
\begin{figure}[htb]
\begin{center}
\mbox{\epsfig{file=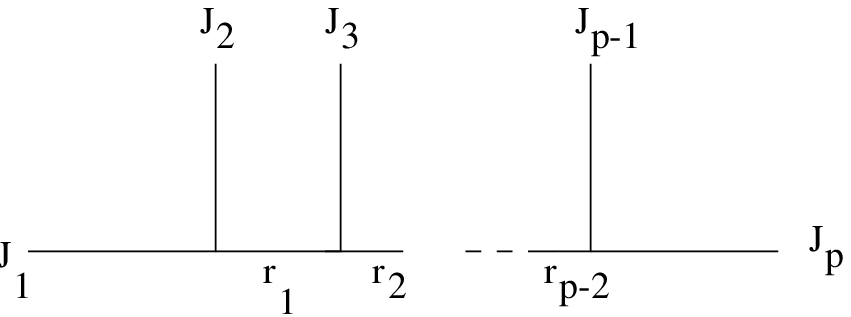,width=16truecm,angle=0}}
\caption{} 
\end{center}
\end{figure}
Here, each matrix element $N_{ij}^{~~r}$ is $1 ~or~ 0$, depending on whether the
primary field  $[\phi_r]$ is allowed  or not in the conformal field theory
fusion algebra for the primary fields $[\phi_i]$ and $[\phi_j] $ ~~($i,j,r~ =~ 0,
1/2, 1, ....k/2$):
\begin{equation}
[\phi_i] ~ \otimes~ [\phi_j]~=~~\sum_r~N_{ij}^{~~r} [\phi_r]~ . \label{fusal}
\end{equation}
Eq. (\ref {fun}) gives the number of conformal blocks with spins $j_1, j_2, \dots,
j_p$ on $p$ external lines and spins $r_1, r_2, \dots, r_{p-2}$ on
the internal lines. 

We next take recourse to the Verlinde formula \cite{dms}
\begin{equation}
N_{ij}^{~~r}~=~\sum_s~{{S_{is} S_{js} S_s^{\dagger r }} \over S_{0s}}~, \label{verl}
\end{equation}
where, the unitary matrix $S_{ij}$ diagonalizes the fusion  matrix. Upon using the 
unitarity of the $S$-matrix, the algebra (\ref{fun}) reduces to 
\begin{equation}
N^{\cal P}~=~ \sum_{r=0}^{k/2}~{{S_{j_1~r} S_{j_2~r} \dots S_{j_p~r}} \over 
(S_{0r})^{p-2}}~. \label{red} \end{equation}
Now, the matrix elements of $S_{ij}$ are known for the case under 
consideration ($SU(2)_k$ Wess-Zumino model); they are given by
\begin{equation}
S_{ij}~=~\sqrt{\frac2{k+2}}~sin \left({{(2i+1)(2j+1) \pi} \over k+2} \right )~, 
\label{smatr} \end{equation}
where, $i,~j$ are the spin labels, $i,~j ~=~ 0, 1/2, 1,  .... k/2$. Using this 
$S$-matrix, the number of conformal blocks for the set of 
punctures ${\cal P}$ is given by
\begin{equation}
N^{\cal P}~=~{2 \over {k+2}}~\sum_{r=0}^{ k/2}~{ {\prod_{l=1}^p sin \left( 
{{(2j_l+1)(2r+1) \pi}\over k+2} \right) } \over {\left[ sin \left( {(2r+1) \pi 
\over k+2} \right)\right]^{p-2} }} ~. \label{enpi} \end{equation}
In the notation of \cite{ash4}, eq. (\ref{enpi}) gives the dimensionality, $dim 
~{\cal 
H}^{\cal P}_S$, {\it for arbitrary area of the horizon $k$ and arbitrary number of 
 punctures}. The dimensionality of the space of states ${\cal H_S}$ of CS theory on 
three-manifold with $S^2$ boundary  is then given  by summing $N^{\cal P}$ over all sets 
of punctures ${\cal P}, ~ N_{bh}~=~\sum_{\cal P} N^{\cal P}$. Then, the entropy of the 
black hole is given by $S_{bh}~=~\log N_{bh}$.

Observe now that eq. (\ref{enpi}) can be rewritten, with appropriate redefinition of 
dummy variables and recognizing that the product can be written as a multiple sum,
\begin{equation}
N^{\cal P}~=~\left ( 2 \over {k+2} \right) ~\sum_{l=1}^{k+1} sin^2 
\theta_l~\sum_{m_1 = 
-j_1}^{j_1} \cdots \sum_{m_p=-j_p}^{j_p} \exp \{ 2i(\sum_{n=1}^p m_n)~ \theta_l \}~, 
\label{summ} \end{equation}
where, $\theta_l ~\equiv~ \pi l /(k+2)$. Expanding the $\sin^2 \theta_l$ and 
interchanging the order of the summations, a few manipulations then yield
\begin{equation}
N^{\cal P}~=~\sum_{m_1= -j_1}^{j_1} \cdots \sum_{m_p=-j_p}^{j_p} \left[ 
~\delta_{(\sum_{n=1}^p m_n), 0}~-~\frac12~ \delta_{(\sum_{n=1}^p m_n), 1}~-~ 
\frac12 ~\delta_{(\sum_{n=1}^p m_n), -1} ~\right ], \label{exct}
\end{equation}
where, we have used the standard resolution of the periodic Kronecker deltas in terms of 
exponentials with period $k+2$,
\begin{equation}
\delta_{(\sum_{n=1}^p m_n), m}~=~ \left( 1 \over {k+2} \right)~ \sum_{l=0}^{k+1} \exp 
\{2i~[ (\sum_{n=1}^p m_n)~-~m] \theta_l \}~. \label{resol}
\end{equation}
Notice that the explicit dependence on $k+2$ is no longer present in the exact formula 
(\ref{exct}).\footnote{A similar method has been advocated in ref. \cite{smo} for 
self-dual boundary conditions}. 

The foregoing calculation does not assume any restrictions on $k$ or number of 
punctures $p$. {\it Eq. (\ref{exct}) is thus an exact formula for the quantum entropy of a 
Schwarzschild  black hole of {\it any} size}, not necessarily large compared to the 
Planck size. However, there is no corresponding calculation of the spectrum of 
the area operator 
for arbitrary-sized surfaces; the formula (\ref{darea}) is valid only for ${\cal 
A}_S \gg 
l_{Planck}^2$. This is what should be delineated as the {\it semiclassical} limit of the 
theory, a domain in which true quantum gravitational effects do not yet make their 
appearance. In this restricted regime, our result (\ref{enpi}) reduces to 
\begin{equation}
N^{\cal P}~~\sim~~\prod_{l=1}^p~(2j_l~+~1)~~\label{bigk}
\end{equation}
in agreement with the result of ref. \cite{ash4}. It is not difficult to show that, upon 
performing the sum over all possible punctures, for large number of punctures and
areas ($k \gg 1$), $S_{bh}={\cal A}_S / 4 G \hbar$  for a specific value of the BI 
parameter \cite{ash4}. To see if the B-H formula relating entropy with area is valid even 
when this restriction of large horizon area is lifted, as it would be in the
full quantum theory, one needs to obtain the eigenvalues of the area operator without any
assumptions about their size. This might entail a modified regularized area operator which
measures horizon area in the quantum theory and is, in general, a constant of motion,
i.e., commutes with the Hamiltonian constraint. The completion of this part of the task 
should reveal quantum corrections to the semi-classical B-H formula. 

The one ambiguity that has remained throughout the calculation of the entropy of
the Schwarzschild black hole is the BI parameter. This was `fixed' by the
requirement that the microstate counting actually gives the area law with the
Hawking value of the proportionality constant. It has been claimed that this choice
is universal for non-rotating black holes \cite{ash4}. Recall that the
existence of this parameter is independent of whether the  spacetime one is quantizing is 
a black hole or not; it is an inherent aspect of canonical quantum gravity.  Therefore, 
it is not unlikely that it will actually be determined (maybe
through an eigenvalue equation) from the Hamiltonian constraint, when one gets a
better understanding of that constraint \cite{imm}. 

It appears that methods of two dimensional conformal field theory effectively
describe {\it quantitative} quantum physics of the black holes in four spacetime
dimensions. In this respect, the similarity with the computational method adopted
in the string theoretic approach, is quite remarkable, although the two conformal
field theories used in the two paths remain quite different.\footnote{Although this
seemed to be the case for the three dimensional BTZ black hole as well \cite{str},
it has now been established \cite{car} that they are indeed the same if looked at
more carefully.} Black hole entropy appears to be a global property that requires a
counting of microstates of the horizon; in the canonical quantum gravity case, this
number is most likely a topological quantity, as it counts the boundary states of a
three dimensional topological field theory. The calculation of the degeneracy of
the `long string' states in the D-brane approach might also share this property. In this 
respect, mention should be made perhaps of the ideas of 't Hooft and Susskind, on 
attempts to model the horizon as a `hologram' \cite{hoof}, \cite{suss}. These ideas were 
vary likely forerunners of the more recent realization that `horizon properties' of 
realistic four dimensional black holes can be computed on the basis of two dimensional 
(conformal) field theories. The commonality that one observes between the two approaches 
with premises that have almost nothing to do with each other is indeed striking. 

There is however one major flaw in both approaches; the black hole curvature 
singularity seems to play no role at all. Thus, the two methods of computing number of 
microstates go through for any non-trivial boundary of spacetime (like 
a horizon), even though there may not be any spacetime singularity beyond that 
boundary.\footnote{See, e.g., ref. \cite{hus} on this issue.} By concentrating only on 
posible properties of the horizon, an essential aspect of black holes is probably being 
ignored. In other words, none of these two approaches seems to illumine, in any manner 
whatsoever, 
the quantum nature of gravitational collapse. It stands to reason that any fundamental 
theory of quantum gravity will have to deal directly with spacetime singularities, just 
as quantum electrodynamics resolves the singularity problems of the Maxwell theory.

\end{document}